\documentclass[conference,letterpaper]{IEEEtran}

\usepackage[utf8]{inputenc}
\usepackage[T1]{fontenc}
\usepackage{url}
\usepackage{ifthen}
\usepackage{cite}
\usepackage[cmex10]{amsmath} 

\usepackage{amsfonts}    
\usepackage{color}
\usepackage{graphicx}
\usepackage[dvips]{epsfig}
\usepackage{graphics}
\usepackage{arydshln}
\usepackage{amsmath} 
\usepackage{amssymb}  
\usepackage{subfigure}
\usepackage{subfloat}
\usepackage{caption}
\usepackage{algorithm}
\usepackage{algorithmic}
\usepackage{hyperxmp}
\usepackage{hyperref}

\def\bP{\mathbb{P}}

\def\cU{\mathcal{U}}

\def \qed {\hfill \vrule height6pt width 6pt depth 0pt}
\def\bee{\begin{equation}}
\def\ene{\end{equation}}
\def\beq{\begin{eqnarray}}
\def\enq{\end{eqnarray}}
\def\been{\begin{equation*}}
\def\enen{\end{equation*}}

\newtheorem{pro}{Proposition}[section]
\newtheorem{lem}{Lemma}[section]

\begin{document}
\title{Dynamic Pricing for Controlling Age of Information}


\author{%
  \IEEEauthorblockN{Xuehe Wang}
  \IEEEauthorblockA{Infocomm Technology Cluster,\\
  Singapore Institute of Technology, Singapore\\
                    Email: xuehe.wang@singaporetech.edu.sg}
  \and
  \IEEEauthorblockN{Lingjie Duan}
  \IEEEauthorblockA{Engineering Systems and Design Pillar,\\
Singapore University of Technology and Design, Singapore\\
Email: lingjie\_duan@sutd.edu.sg}
}

\maketitle

\begin{abstract}

Fueled by the rapid development of communication networks and sensors in portable devices, today many mobile users are invited by content providers to sense and send back real-time useful information (e.g., traffic observations and sensor data) to keep the freshness of the providers' content updates. However, due to the sampling cost in sensing and transmission, an individual may not have the incentive to contribute the real-time information to help a content provider reduce the age of information (AoI). Accordingly, we propose dynamic pricing for the provider to offer age-dependent monetary returns and encourage users to sample information at different rates over time. This dynamic pricing design problem needs to balance the monetary payments to users and the AoI evolution over time, and is challenging to solve especially under the incomplete information about users' arrivals and their private sampling costs. For analysis tractability, we linearize the nonlinear AoI evolution in the constrained dynamic programming problem, by approximating the dynamic AoI reduction as a time-average term and solving the approximate dynamic pricing in closed-form. Then, we estimate this approximate term based on Brouwer's fixed-point theorem. Finally, we provide the steady-state analysis of the optimized approximate dynamic pricing scheme for an infinite time horizon, and show that the pricing scheme can be further simplified to an $\varepsilon$-optimal version without recursive computing over time.

\end{abstract}



\section{Introduction}

Many customers today prefer not to miss any useful information or breaking news even if in minute, making it imperative for a content provider to keep the posted information fresh to attract a good number of customers for profit \cite{fiveindustries},\cite{google}. The real-time information can be traffic condition, news, sales promotion, and air quality index, and they will gradually become outdated and useless over time. To keep information fresh, many content providers such as Waze and CrowdSpark now invite and pay the mobile crowd including smartphone users and drivers to sample real-time information frequently \cite{duan2014motivating}. Such crowdsensing approach also avoids a content provider's own deployment of an expensive fixed sensor network across the city or nation. The fast development of wireless communication networks and sensors in portable devices enables the mobile users to contribute real-time information. 

Age of information (AoI) is recently proposed as an important performance metric to quantify the freshness of the information in these online applications. The literature focuses on the technological issues of the AoI such as the frequency of status updates and queueing delay analysis. In \cite{kaul2012real}, the communication time of the status update systems is considered, and it proves the existence of an optimal packet generation rate at a source to keep its status as timely as possible. Noting the time-varying availability of energy at the source will affect the update packet transmission rate, \cite{bacinoglu2015age} derives an offline solution that minimizes both the time average age and the peak age for an arbitrary energy replenishment. \cite{sun2017update} shows a counter-intuitive phenomenon that zero-wait policy, i.e., a fresh update is submitted once the previous update is delivered, does not always minimize the age. Considering random packet arrivals, \cite{hsu2017age} studies how to keep many customers updated over a wireless broadcast network and a Markov decision process (MDP) is formulated to find dynamic scheduling algorithms. For the analysis tractability, the Peak Age-of-Information (PAoI) metric, which is the average maximum age before a new update is received, was first considered in \cite{costa2014age} for a single-class M/M/1 queueing system. 


However, the economic issues of controlling AoI for content providers are largely overlooked in the literature. On one hand, individuals incur sampling costs when sense and send back their real-time information to content providers, and they should be rewarded and well motivated to contribute their information updates \cite{duan2014motivating}. On the other hand, a large crowdsensing pool implies a large total sampling cost to compensate, which should be taken into account in a content provider's sustainable management of its AoI \cite{shugangAOI2019}, \cite{zhangmeng2019}. As AoI changes over time, the pricing should be dynamic and age-dependent to best balance the AoI evolution and the sampling cost to compensate, yet this dynamic programming problem is usually difficult to solve due to the curse of dimensionality. Further, we face another challenge for optimally deciding a provider's dynamic pricing: incomplete information about users' private sampling costs and their random arrival to help sample. Individuals are different in nature and incur different sampling costs to reflect their heterogeneity (e.g., in battery energy storage and privacy concern when sampling). A user will accept the price if his sampling cost is less than the price offered by the provider, yet the provider does not know such private cost when deciding pricing. In addition, users are mobile and their arrivals in the target area to sense is random. 

To our best knowledge, this paper is the first work studying the dynamic pricing issue for controlling a content provider's AoI. Under incomplete information, we study how the provider should decide its dynamic pricing to minimize the discounted AoI and monetary payment over time, by taking into account the private sampling costs of users and random user arrival in the target area. After formulating this problem as a nonlinear dynamic program in Section \ref{sec_systemmodel}, for analysis tractability, we linearize the nonlinear AoI evolution in the constrained dynamic programming problem in Section \ref{sec_approximate_dynamic}, by using a time-average term to estimate the dynamic AoI reduction in approximate sense and successfully solving the dynamic pricing scheme in closed-form. Then we determine the time-average estimator, based on Brouwer's fixed-point theorem. Section \ref{sec_steady_infty} further provides the steady-state analysis of the approximate dynamic pricing scheme for an infinite time horizon. It is shown that the approximate dynamic pricing can be further simplified to an $\varepsilon$-optimal version without recursive computing over time. Finally, Section \ref{sec_conclusion} draws the conclusion of the paper.


\section{System Model and Problem Formulation}\label{sec_systemmodel}

\begin{figure}
\centering\includegraphics[scale=0.32]{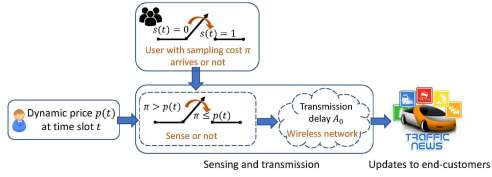}\caption{Illustration of information update process under the content provider's dynamic pricing and random user arrival.}\label{fig_systemmodel}
\end{figure}

We illustrate the information update process at the content provider side in Fig. \ref{fig_systemmodel}. We consider a discrete time horizon with time slot $t=0,1,\cdots$. The provider first announces price $p(t)$ at the beginning of time slot $t$, and a user may arrive randomly in this time slot and (if so) he further decides to sample or not based on the price $p(t)$ and its own sampling cost $\pi$. If the user appears and accepts to sample ($\pi\leq p(t)$), its sensor data (e.g., about traffic and road condition) is transmitted with fixed delay $A_0$ to finally reach the end customers (who use the content provider's app).

As in \cite{hsu2017age}, we consider that the users' random arrivals in the target area (to the content provider's interest) are independent and identically distributed (i.i.d.) over time slots, by following a Bernoulli distribution. As shown in Fig. \ref{fig_systemmodel}, if a user arrives in time slot $t$, $s(t)=1$; otherwise, $s(t)=0$, where the probability of one user arrival in each time slot is $\alpha$, i.e., $\bP(s(t)=1)=\alpha$. Each time slot's duration is properly selected such that there is at most one user arrival at a time. Further, the users' sampling costs are i.i.d. according to a cumulative distribution function (CDF) $F(\pi), \pi\in[0,b]$. Though all potential users' costs follow the same distribution, their realized costs are different in general. Under the incomplete information, the provider does not know the users' arrivals for potential sampling over time $t$ or the arriving user's particular cost $\pi$. It only knows the user arrival probability $\alpha$ in each time slot and the cost distribution $F(\pi)$.


\begin{figure}
\centering\includegraphics[scale=0.33]{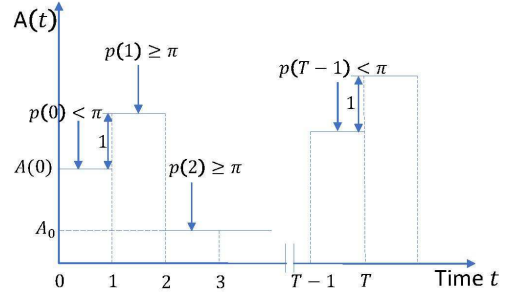}\caption{Actual age $A(t)$ over time under dynamic pricing $p(t)$.}\label{fig_aoimodeltra}
\end{figure}

We adopt the Age of Information (AoI) as the performance metric to quantify the freshness of the information packet at end-customer side. Let $A(t)$ be the AoI at the beginning of time slot $t$. Considering a linearly increasing actual age over time as in Fig. \ref{fig_aoimodeltra} for the discrete time horizon, the new age $A(t+1)$ at time $t+1$ increases by one, i.e., $A(t)+1$, if the information is not updated by any user at time $t$. If a user arrives in time $t$ and further accepts the price $p(t)$, i.e., $\pi\leq p(t)$, a new status packet will be generated and transmitted. Without much loss of generality, we assume the status sampling and transmission are accomplished within a time slot,\footnote{This is feasible provided with the upcoming ultra-reliable and low-latency 5G communications.} then the transmission delay is fixed to $A_0\leq 1$ and the age $A(t+1)$ at time $t+1$ decreases to $A_0$ if an update is received. Then, the dynamics of the actual AoI is given as
\begin{equation}
A(t+1)=\left\{
\begin{array}{l}
A_0, ~~~~~~~\text{if $\pi\leq p(t)$;} \\
A(t)+1,  \text{otherwise.}\\
\end{array}
\right.
\end{equation}


\begin{figure*}[t]
\centering
\subfigure[Approximate dynamic pricing $p(t)$ versus time $t$.]{\label{subfig_p(t)}
\begin{minipage}{.23\textwidth}
\includegraphics[width=1\textwidth]{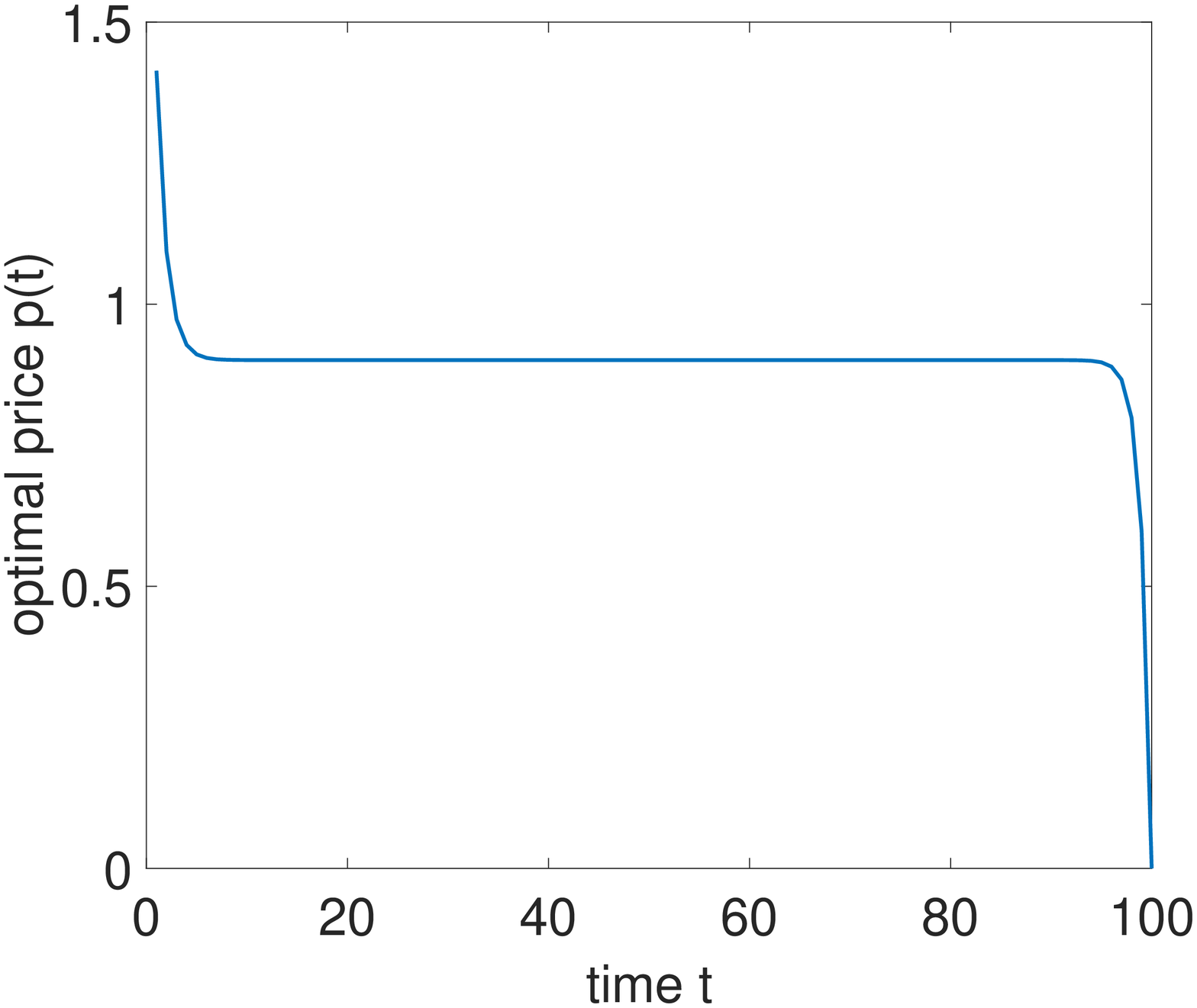}
\end{minipage}
}
\subfigure[Expected age $A(t)$ versus time $t$.]{\label{subfig_age}
\begin{minipage}{.23\textwidth}
\includegraphics[width=1\textwidth]{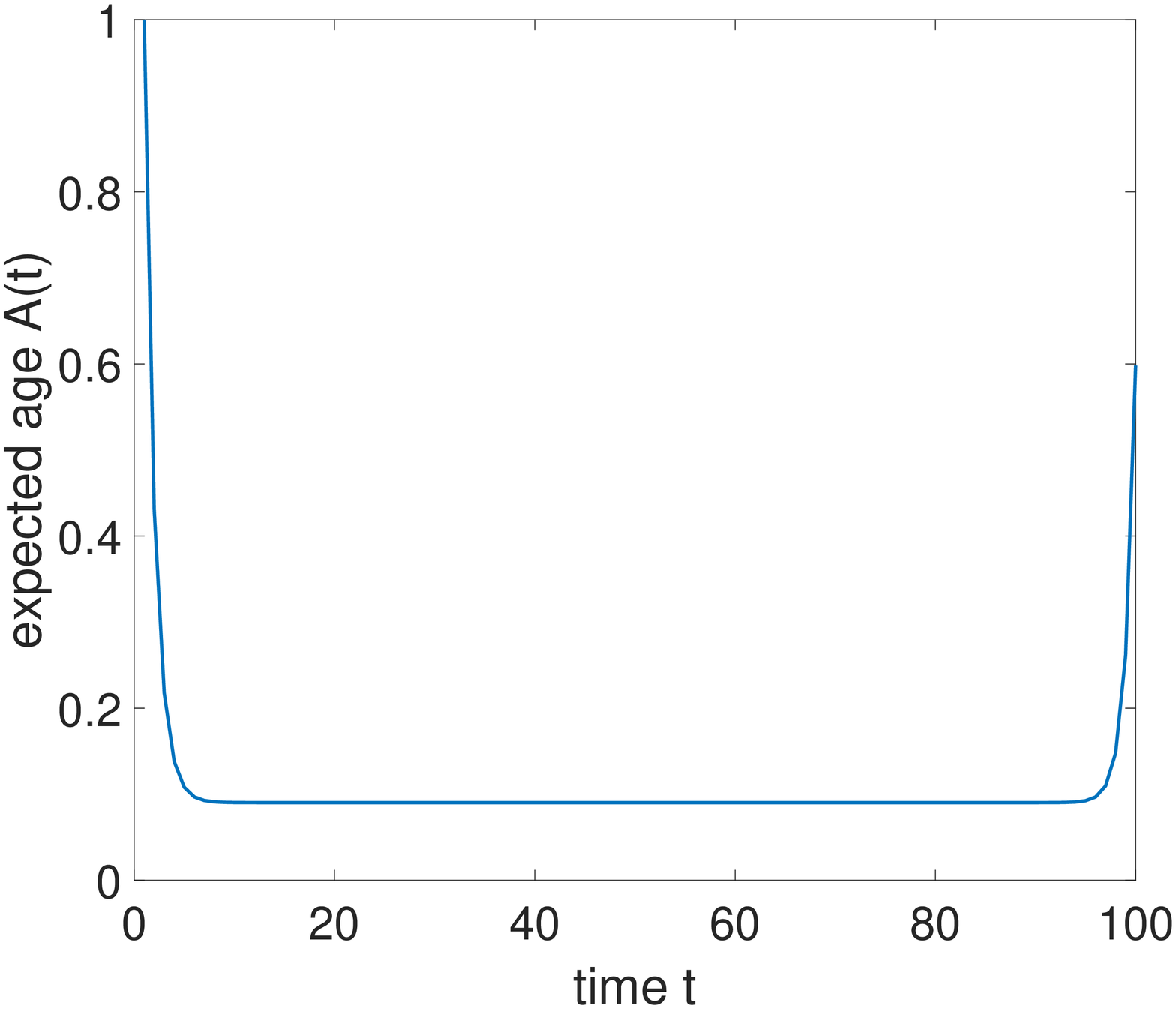}
\end{minipage}
}
\subfigure[$Q_t$ versus time $t$.]{\label{subfig_Qt}
\begin{minipage}{.23\textwidth}
\includegraphics[width=1\textwidth]{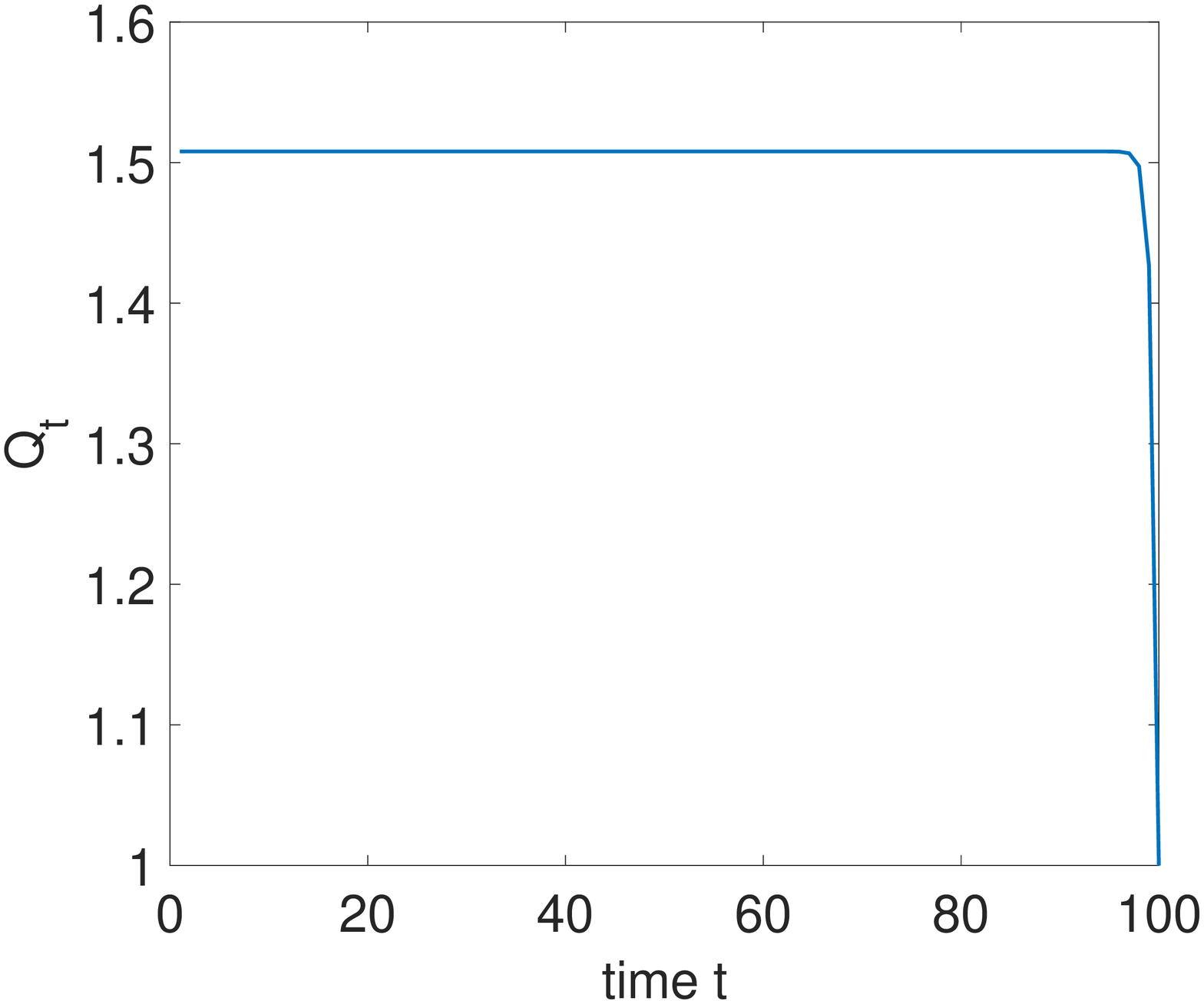}
\end{minipage}
}
\subfigure[$M_t$ versus time $t$.]{\label{subfig_Mt}
\begin{minipage}{.23\textwidth}
\includegraphics[width=1\textwidth]{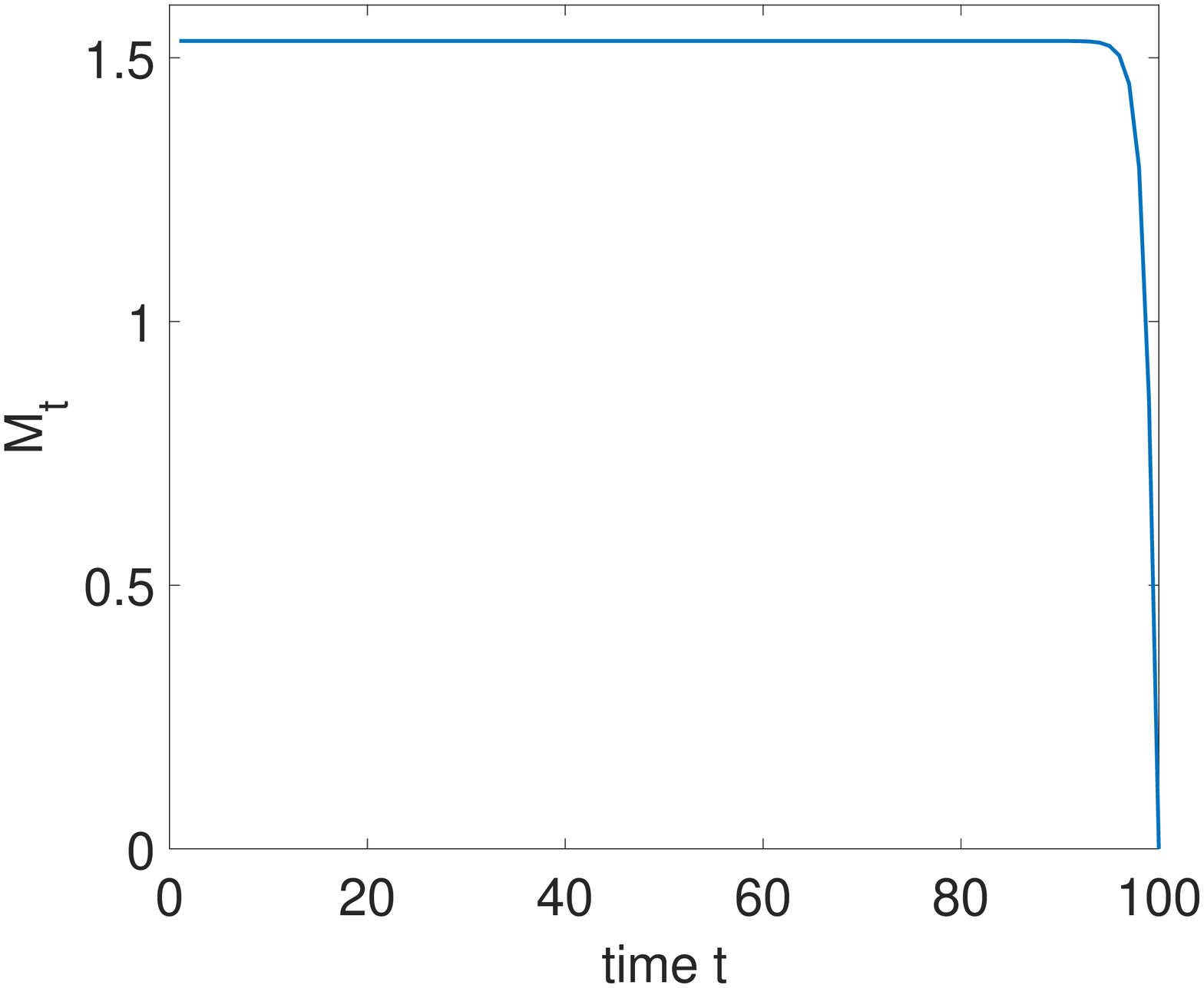}
\end{minipage}
}
\caption{Dynamics of price $p(t)$, expected age $A(t)$, $Q_t$, and $M_t$ over time for finite horizon $T=100$.}\label{fig_dynamics}
\end{figure*}

Given the cumulative distribution function of an arriving user's cost $F(\pi)$, the probability that a user will appear and accept the price at time $t$ is $\alpha F(p(t))$. Considering uniform distribution for the users' private costs (i.e., $F(\pi)=\frac{\pi}{b}$, $\pi\in [0,b]$), the expected age at time $t+1$ is
\bee\begin{split}\label{equ_A_dynamic} A(t+1)=&A_0\alpha F(p(t))+(A(t)+1)(1-\alpha F(p(t)))\\
=&A(t)-(A(t)-A_0)\alpha \frac{p(t)}{b}+(1-\alpha\frac{p(t)}{b}), \end{split}\ene
which tells nonlinear AoI evolution due to the product between $A(t)$ and $p(t)$ above.

Since the probability that a user appears and accepts the price $p(t)$ is $\frac{\alpha p(t)}{b}$, the expected payment to the user is $\frac{\alpha p^2(t)}{b}$. Note that the optimal price $p(t)$ should not exceed the maximum cost $b$ of the user. The objective of the provider is to find the optimal dynamic pricing $p(t), t\in[0,...,T]$ that minimizes the expected total discounted cost, which is the summation of the square age and monetary payment in expected sense:
\bee\label{equ_UTobjective} U(T)=\min_{p(t)\in [0,b],t\in[0,...,T]}\sum_{t=0}^{T}\rho^t(A^2(t)+\frac{\alpha p^2(t)}{b}), \ene
\been \text{s.t.}~~A(t+1)=A(t)-(A(t)-A_0)\alpha \frac{p(t)}{b}+(1-\alpha\frac{p(t)}{b}), ~(\ref{equ_A_dynamic})\enen
where $\rho\in(0,1)$ is the discounted factor and we choose the square age $A^2(t)$ to reflect the fact that the provider's profit loss should be convexly increasing in its age.

The problem in (\ref{equ_A_dynamic})-(\ref{equ_UTobjective}) is a constrained nonlinear dynamic program, which is challenging to solve analytically due to the curse of dimensionality. We can imagine that, a huge number of combinations of the time-dependent prices $p(t), t\in[0,...,T]$ must be jointly designed, and the computation complexity $O(b^T)$ is formidably high by increasing exponentially in $T$. To analytically solve this problem for obtaining useful insights, we will propose a problem approximation in the following section.

\section{Approximation of Dynamic Pricing}\label{sec_approximate_dynamic}

To analytically obtain the optimal dynamic prices $p(t), t\in[0,...,T]$ for the constrained dynamic problem in (\ref{equ_UTobjective}), we reformulate the nonlinear dynamics of the expected age in (\ref{equ_A_dynamic}) into linear dynamics. Specifically, we propose a time-average term $\delta$ as an estimator to approximate the dynamic age reduction $A(t)-A_0$ per update in (\ref{equ_A_dynamic}), i.e.,
\bee\label{equ_A_dynamic_estimate} A(t+1)=A(t)-\delta\alpha \frac{p(t)}{b}+(1-\alpha \frac{p(t)}{b}), \ene
where the time-average estimator $\delta$ also takes into account the time discount factor $\rho$, i.e.,\footnote{There is no need to include $A(T)$ in $\delta$, as it will not affect the AoI $A(t)$ at previous time slot $t\leq T-1$.}
\bee\label{equ_delta} \delta=\frac{1-\rho}{1-\rho^T}\sum_{t=0}^{T-1}\rho^t (A(t)-A_0). \ene


In the following, we first assume the estimator $\delta$ is known for analyzing the approximate dynamic pricing in Section \ref{sec_approx_pricing1}. Later in Section \ref{sec_fixed_delta}, we will show how to determine the estimator $\delta$ for pricing update.

\subsection{Analysis of approximate dynamic pricing}\label{sec_approx_pricing1}

Though the dynamic programming problem (\ref{equ_UTobjective})-(\ref{equ_A_dynamic_estimate}) now has only linear AoI evolution constraint in (\ref{equ_A_dynamic_estimate}), it is still not easy to solve by considering the huge number of price combinations over time. We denote the cost function with initial time $t$ as
\bee\label{equ_J_from_t} \begin{split} J(p,t)
=\sum_{s=t}^{T}\rho^{s-t}(A^2(t)+\frac{\alpha}{b}p^2(t)),
\end{split}\ene
and denote the value function as
\bee\label{equ_Vi} V(A(t),t)=\min_{\{p(s)\in [0,b]\}_{s=t}^T}(J(p,t)|A(t)). \ene
Then, we have the dynamic programming equation:
{\small \bee\label{equ_VAt}\begin{split} V(A(t),t)=&\min_{p(t)\in [0,b]}(A^2(t)+\frac{\alpha}{b}p^2(t)+\rho V(A(t+1),t+1)),\end{split}\ene}
subject to (\ref{equ_A_dynamic_estimate}).

In the following, we first analyze the unconstrained dynamic problem (\ref{equ_VAt}) by using dynamic control techniques. Later in Section \ref{sec_steady_infty}, the constraint $p(t)\in[0,b]$ will be added back and examined. 


\begin{pro}\label{pro_optimal_pt} The approximate dynamic pricing $p(t), t\in\{0,...,T\}$ as an optimal solution to the unconstrained dynamic program (\ref{equ_VAt}) is increasing in $A(t)$ and given by
\bee\label{equ_pt_finite} p(t)=\frac{\rho M_{t+1}(\delta+1)+2\rho(\delta+1)Q_{t+1}(A(t)+1)}{2+2\rho Q_{t+1}\frac{\alpha(\delta+1)^2}{b}}, \ene
and the resulting expected age $A(t)$ at time $t$ is
\bee\label{equ_At_non_infty}\begin{split} A(t)=&\prod_{i=1}^t\frac{1}{1+\rho Q_i\frac{\alpha(\delta+1)^2}{b}}A(0)+\frac{2-\rho M_t\frac{\alpha(\delta+1)^2}{b}}{2+2\rho Q_t\frac{\alpha(\delta+1)^2}{b}}\\
&+\sum_{s=1}^{t-1}\frac{2-\rho M_s\frac{\alpha(\delta+1)^2}{b}}{2+2\rho Q_s\frac{\alpha(\delta+1)^2}{b}}\prod_{i=s+1}^t\frac{1}{1+\rho Q_i\frac{\alpha(\delta+1)^2}{b}}, \end{split}\ene
where
\bee\label{equ_Qt} Q_t=1+\frac{\rho Q_{t+1}}{1+\rho Q_{t+1}\frac{\alpha(\delta+1)^2}{b}}, \ene
\bee\label{equ_Mt} M_t=\frac{\rho(M_{t+1}+2 Q_{t+1})}{1+\rho Q_{t+1}\frac{\alpha(\delta+1)^2}{b}}, \ene
with $p(T)=0, Q_T=1, M_T=0$. Fig. \ref{fig_dynamics} illustrates their dynamics over time.
\end{pro}

\textbf{Proof Sketch:} According to $\frac{\partial V(A(t),t)}{\partial p(t)}=0$, we can observe that $p(t)$ is a linear function of $A(t)$. Thus, the value function should be in the following quadratic structure:
\bee\label{equ_VAguess} V(A(t),t)=Q_tA^2(t)+M_tA(t)+S_t, \ene
yet we still need to determine $Q_t, M_t, S_t$. This will be accomplished by finding the recursion in the following.

First, we have $Q_T=1, M_T=0, S_T=0$ due to $V(A(T),T)=A^2(T)$. Given $V(A(t+1),t+1)=Q_{t+1}A^2(t+1)+M_{t+1}A(t+1)+S_{t+1}$ as in (\ref{equ_VAguess}), the dynamic programming equation at time $t$ is
\bee\label{equ_Vt}\begin{split} V(A(t),t)=&\min_{p(t)}\Big(A^2(t)+\frac{\alpha}{b}p^2(t)+\rho Q_{t+1}A^2(t+1)\\
&+\rho M_{t+1}A(t+1)+\rho S_{t+1}\Big).\end{split}\ene

Insert (\ref{equ_A_dynamic_estimate}) into (\ref{equ_Vt}) and let $\frac{\partial V(A(t),t)}{\partial p(t)}=0$, we have the optimal price $p(t)$ as given in (\ref{equ_pt_finite}). Then, insert $p(t)$ in (\ref{equ_pt_finite}) into $V(A(t),t)$ in (\ref{equ_Vt}), we have $V(A(t),t)$ as a function of $Q_{t+1}, M_{t+1}, S_{t+1}$ and $A(t)$. Thus, by reformulating $V(A(t),t)$ in (\ref{equ_Vt}) and noting that $V(A(t),t)=Q_tA^2(t)+M_tA(t)+S_t$, we obtain the recursive relationships in (\ref{equ_Qt}) and (\ref{equ_Mt}). Insert  $p(t)$ in (\ref{equ_pt_finite}) into (\ref{equ_A_dynamic_estimate}), we obtain the expected age $A(t)$ in (\ref{equ_At_non_infty}). \qed

In the simulation results of Fig. \ref{fig_dynamics}, we can see that the dynamic pricing $p(t)$ first decreases with the expected age $A(t)$ until both of them reach steady-states, which is consistent with the proportional relationship between $p(t)$ and $A(t)$ in (\ref{equ_pt_finite}). But when close to the end of the time horizon $T$, the price $p(t)$ decreases to $0$ to save sampling expense without worrying its negative effect on the age. The expected age $A(t)$ increases again but only lasts for a few time slots. Noting that $Q_t$ and $M_t$ are computed in backward recursion over time, we can see that both $Q_t$ and $M_t$ fast converge, which will be strictly proved in Section \ref{sec_steady_infty}.

\subsection{Update of estimator $\delta$ for pricing}\label{sec_fixed_delta}


Now we are ready to update the estimator $\delta$ (which approximately linearizes (\ref{equ_A_dynamic}) as (\ref{equ_A_dynamic_estimate})) for dynamic pricing in (\ref{equ_pt_finite}) and find the fixed point to fit (\ref{equ_delta}). We should also note that this fixed point may or may not exist. Note that the estimator $\delta$ in (\ref{equ_delta}) is affected by all the ages $A(t)$ over the time horizon $t\in\{0,...,T-1\}$, which will in turn affect $A(t)$. Thus, according to (\ref{equ_At_non_infty}), $A(t)$ is a function of all the ages $\{A(t), t\in\{0,...,T-1\}\}$ over time and we need to find the estimator $\delta$ such that it replicates $\frac{1-\rho}{1-\rho^T}\sum_{t=0}^{T-1}\rho^t (A(t)-A_0)$ in (\ref{equ_delta}) as we initially assumed.

For any $1\leq t \leq T-1$, we insert $\delta=\frac{1-\rho}{1-\rho^T}\sum_{t=0}^{T-1}\rho^t (A(t)-A_0)$ into $A(t)$ in (\ref{equ_At_non_infty}), and define
\bee\label{equ_phitA}\begin{split} &\Phi_t(A(1),\cdots,A(T-1))=A(t).  \end{split}\ene
Then, we have the following vector function for $t\in[1,T-1]$:
{\small \begin{align} &\Phi(A(1),\cdots,A(T-1))\\=&(\Phi_1(A(1),\cdots,A(T-1)),\cdots,\Phi_{T-1}(A(1),\cdots,A(T-1))).\notag \end{align}}
According to (\ref{equ_phitA}), the fixed point in $\Phi(A(1),...,A(T-1))=(A(1),...,A(T-1))$ should be reached to tell that $\delta$ replicates $\frac{1-\rho}{1-\rho^T}\sum_{t=0}^{T-1}\rho^t (A(t)-A_0)$.


Note that $Q_t\geq 1$ in (\ref{equ_Qt}), $M_t\geq 0$ in (\ref{equ_Mt}), we have \bee \Phi_t\leq A(0)+t. \ene Define $\Omega=[0,A(0)+1]\times\cdots\times[0,A(0)+(T-1)]$. Since $\Phi_t$ is continuous in $\Omega$, $\Phi$ is a continuous mapping from $\Omega$ to $\Omega$. According to the Brouwer's fixed-point theorem, we have the following proposition.


\begin{pro} $\Phi$ has a fixed point in $\Omega$.
\end{pro}

Given the existence of the fixed point, we are ready to find the estimator $\delta$ that is consistent with our approximation assumption in (\ref{equ_delta}). Accordingly, we propose Algorithm \ref{alg_find_fixed_b1}: given any initial estimator $\delta^{est}(j)$ in round $j$, we can iteratively obtain the resulting expected ages $A(t), t\in[1,T-1]$ according to (\ref{equ_At_non_infty}), and then check whether the resulting estimator $\delta^{est}(j+1)$ in next round coincides with the initial estimator $\delta^{est}(j)$. By repeating the process until $\delta^{est}(j+1)=\delta^{est}(j)$, we obtain the fixed point $\delta$ and the computation complexity of Algorithm \ref{alg_find_fixed_b1} is $O(\frac{T}{\epsilon})$.

\begin{algorithm}[t]
\caption{Iterative computation of fixed point estimator $\delta$}
\begin{algorithmic}[1]

\STATE $\epsilon=1,j=1$, an arbitary initial $\delta^{est}(0)\geq 0$, $\delta=\delta^{est}(0)$

\WHILE {$\epsilon>0.001$}

\FOR {$t=0$ to $T-1$}
\STATE Compute $Q_t$ and $M_t$ according to $\delta$, (\ref{equ_Qt}), (\ref{equ_Mt})
\ENDFOR

\FOR {$t=1$ to $T-1$}
\STATE Compute $A(t)$ according to (\ref{equ_At_non_infty})
\ENDFOR

\STATE $\delta^{est}(j)=\frac{1-\rho}{1-\rho^T}\sum_{t=0}^{T-1}\rho^t (A(t)-A_0)$
\STATE $\delta=\delta^{est}(j)$
\STATE $\epsilon=\delta^{est}(j)-\delta^{est}(j-1)$
\STATE $j=j+1$

\ENDWHILE
\RETURN Fixed point $\delta$

\end{algorithmic}
\label{alg_find_fixed_b1}
\end{algorithm}

\section{Steady-state Analysis of Dynamic Pricing}\label{sec_steady_infty}


We wonder how our approximate dynamic pricing and its performance would be in the steady-state, by looking at the infinite time horizon $T\rightarrow\infty$ in this section. Specifically, the steady-state characterizations of $Q_t$ in (\ref{equ_Qt}) and $M_t$ in (\ref{equ_Mt}) can be found by iterating the dynamic equations until they converge. The following lemma shows the steady-states of $Q_t$ and $M_t$, both of which exist and are nicely given in closed-form.

\begin{lem}\label{lem_Mt_Q_t_infinity} As $T\rightarrow\infty$, $Q_t$ and $M_t$ respectively converge to the steady-states: {\small \bee\label{equ_solve_Q} Q=\frac{1}{2}\Big(1-\frac{b(1-\rho)}{\rho \alpha(\delta+1)^2}+\sqrt{(1-\frac{b(1-\rho)}{\rho \alpha(\delta+1)^2})^2+\frac{4b}{\rho\alpha(\delta+1)^2}}\Big), \ene
\bee\label{equ_stableM} M=\frac{2\rho Q}{1-\rho+\rho Q\frac{\alpha(\delta+1)^2}{b}}. \ene}
\end{lem}

\textbf{Proof:} Since both $Q_{t}=1+\frac{\rho Q_{t+1}}{1+\rho Q_{t+1}\frac{\alpha(\delta+1)^2}{b}}$ in (\ref{equ_Qt}) and $M_t=\frac{\rho(M_{t+1}+2 Q_{t+1})}{1+\rho Q_{t+1}\frac{\alpha(\delta+1)^2}{b}}$ in (\ref{equ_Mt}) increase with $Q_{t+1}$ and $M_{t+1}$, respectively, we can conclude that $\{Q_T, Q_{T-1}, ...\}$ and $\{M_T, M_{T-1}, ...\}$ are increasing sequences and converge to the steady state $Q$ and $M$, respectively. By removing the time subscripts from (\ref{equ_Qt}) and (\ref{equ_Mt}), we can show the steady-state $Q$ and $M$ exist and are given in (\ref{equ_solve_Q}) and (\ref{equ_stableM}), respectively. \qed

For the infinity horizon case, the optimal price changes from (\ref{equ_pt_finite}) to
\bee\label{equ_pt_infinity_horizon} p^{\infty}(t)=\frac{\rho M(\delta+1)+2\rho(\delta+1)Q(A(t)+1)}{2+2\rho Q\frac{\alpha(\delta+1)^2}{b}}, \ene
where $Q$ and $M$ are given in (\ref{equ_solve_Q}) and (\ref{equ_stableM}), respectively.

Then, according to the dynamic AoI evolution in (\ref{equ_A_dynamic_estimate}), we have
\bee\label{equ_At_forinfty}\begin{split} A^{\infty}(t)=&(\frac{1}{1+\rho Q\frac{\alpha(\delta+1)^2}{b}})^tA(0)\\
&+\frac{2-\rho M\frac{\alpha(\delta+1)^2}{b}}{2+2\rho Q\frac{\alpha(\delta+1)^2}{b}}\frac{1-(\frac{1}{1+\rho Q\frac{\alpha(\delta+1)^2}{b}})^t}{1-\frac{1}{1+\rho Q\frac{\alpha(\delta+1)^2}{b}}}. \end{split}\ene

By noting that $\frac{1}{1+\rho Q\frac{\alpha(\delta+1)^2}{b}}<1$ and $M$ in (\ref{equ_stableM}), we have the following proposition.

\begin{pro} As $t\rightarrow\infty$, the expected AoI is given as:
\bee\label{equ_At_infty}\begin{split} \lim_{t\rightarrow\infty}A^{\infty}(t)=\frac{(1-\rho)(1+\rho Q\frac{\alpha(\delta+1)^2}{b})}{\rho Q(\delta+1)^2(\frac{\alpha}{b}(1-\rho)+\rho Q(\frac{\alpha(\delta+1)}{b})^2)}, \end{split}\ene
and the optimal dynamic price converges to
\bee\label{equ_pt_infty} \lim_{t\rightarrow\infty}p^{\infty}(t)=\frac{b}{\alpha(\delta+1)}. \ene
\end{pro}


In the following, we will show how to analytically find the fixed point estimator $\delta$ of AoI reduction in infinite horizon and recall that we can only numerically compute it by Algorithm \ref{alg_find_fixed_b1} in Section \ref{sec_fixed_delta} for finite time horizon. Note that $\delta$ is estimated by $\frac{1-\rho}{1-\rho^T}\sum_{t=0}^{T-1}\rho^t (A^{\infty}(t)-A_0)$. Since $A^{\infty}(t)$ converges to the constant in (\ref{equ_At_infty}), as $T\rightarrow\infty$, we can solve the estimator $\delta$ as the unique solution to
\bee\label{equ_solve_b1}\begin{split} \frac{(1-\rho)(1+\rho Q\frac{\alpha(\delta+1)^2}{b})}{\rho Q(\delta+1)^2(\frac{\alpha}{b}(1-\rho)+\rho Q(\frac{\alpha(\delta+1)}{b})^2)}-A_0-\delta=0, \end{split}\ene
which describes the time-average AoI reduction due to user sampling and only non-negative $\delta$ solution makes sense. According to (\ref{equ_solve_b1}), we can obtain the fixed point $\delta$ without using Algorithm \ref{alg_find_fixed_b1} iteratively.



Then, we also add back the constraint $p(t)\in[0,b]$ in dynamic program (\ref{equ_VAt}) to examine, which is relaxed in previous Section \ref{sec_approx_pricing1} for simplifying our analysis, and have the following proposition. 

\begin{pro}\label{pro_bar_c_condition} Both $\delta\geq 0$ and $\lim_{t\rightarrow\infty}p(t)\in[0,b]$ are satisfied to be reasonable if
\bee\label{equ_condition2} \frac{2b(1-\rho)}{\alpha \rho\Big(1-\frac{b(1-\rho)}{\rho \alpha}+\sqrt{(1-\frac{b(1-\rho)}{\rho \alpha})^2+\frac{4b}{\rho\alpha}}\Big)}\geq A_0, \alpha\geq \frac{1}{\delta+1}. \ene
\end{pro}

\textbf{Proof:} By checking the monotonic property of (\ref{equ_solve_b1}) regarding $\delta$, we can prove the uniqueness of the fixed point estimator $\delta$. Then, we show the the condition for $\delta\geq 0$. Rewrite (\ref{equ_solve_b1}) as
\bee\label{equ_solve_b12} \frac{\alpha}{b}=\frac{\rho Q(\frac{\alpha(\delta+1)}{b})^2(\delta+A_0)}{1-\rho}(1-\frac{\rho}{1+\rho Q\frac{\alpha(\delta+1)^2}{b}}). \ene
Denote the right-hand side of (\ref{equ_solve_b12}) as $v(\delta)$. We can check $v(\delta)$ increases with $\delta$. Thus, if $\frac{\alpha}{b}\geq v(\delta=0)$, we have $\delta\geq 0$. Since $v(\delta=0)<\frac{\rho Q(\frac{\alpha}{b})^2(\delta+A_0)}{1-\rho}$, thus we only need $\frac{\alpha}{b}\geq \frac{\rho Q(\frac{\alpha}{b})^2(\delta+A_0)}{1-\rho}$ to satisfy, which is rewritten as the first inequation in (\ref{equ_condition2}). According to (\ref{equ_pt_infty}), $\lim_{t\rightarrow\infty}p^{\infty}(t)\leq b$ always holds given $\alpha\geq \frac{1}{\delta+1}$ and $\delta\geq 0$.  \qed

The condition for reasonable $\delta$ and $p(t)$ in (\ref{equ_condition2}) is likely to hold in many cases. For example, if the transmission delay $A_0$ is small in a crowded area with large $\alpha$, $\lim_{t\rightarrow\infty}p(t)\in[0,b]$ always holds. Actually, under the conditions in (\ref{equ_condition2}), we can safely use the dynamic pricing $p(t)$ in (\ref{equ_pt_finite}) for finite horizon case. If $p(t)>b$ at the first few time slots due to high initial age $A(0)$, we can set $p(t)=b$ to ensure the users (if arrive) to contribute and thus the expected age $A(t)$ will decrease until $p(t)\leq b$. Then, from that time on, $p(t)\in[0,b]$ in (\ref{equ_pt_finite}) is always satisfied under the conditions in (\ref{equ_condition2}). 

\subsection{$\varepsilon$-optimality for Expected Discounted Cost}

\begin{figure}
\centering\includegraphics[scale=0.24]{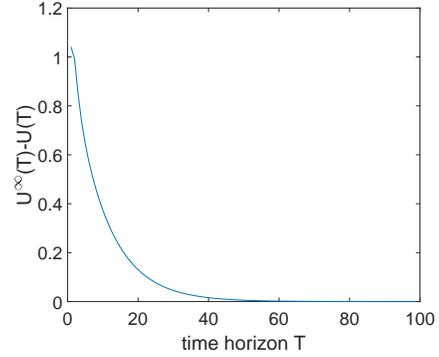}\caption{$U^{\infty}(T)-U(T)$ versus time horizon $T$.}\label{fig_diffU}
\end{figure}


We note that the approximate dynamic pricing is further simplified to (\ref{equ_pt_infinity_horizon}) with the lowest possible computation order $O(1)$ by using the steady-states $Q$ in (\ref{equ_solve_Q}) and $M$ in (\ref{equ_stableM}) for infinite time horizon. It is unlike (\ref{equ_pt_finite}) which still recursively updates $Q_t$ in (\ref{equ_Qt}) and $M_t$ in (\ref{equ_Mt}) for finite horizon case. By using this simple dynamic price $p^{\infty}(t)$ without recursive computing over time, we wonder its performance and denote the resulting expected discounted cost for finite horizon $T$ as $U^{\infty}(T)$. According to Lemma \ref{lem_Mt_Q_t_infinity} and Squeeze Theorem, in the following proposition, we prove that $U^{\infty}(T)$ is $\varepsilon$-optimal compared with the expected discounted cost $U(T)$ under approximate dynamic pricing $p(t)$ in (\ref{equ_pt_finite}). 

\begin{pro} $\forall$ $T>0$, there always exists an $\varepsilon_T>0$ such that
\bee U(T)\leq U^{\infty}(T)\leq U(T)+\varepsilon_T, \ene
and we have $\lim_{T\rightarrow\infty}\varepsilon_T=0$ for a sufficiently large $T$.
\end{pro}

\textbf{Proof:} Since $Q_t, M_t$ converge to the steady state $Q, M$, respectively, there exists a $t_0$ such that for any $t\leq T-t_0$, $Q_t=Q$ and $M_t=M$. Then, for any $t\leq T-t_0$, we have $A(t)=A^{\infty}(t)$. Moreover, according to (\ref{equ_pt_finite}) and (\ref{equ_pt_infinity_horizon}), we have $p(t)=p^{\infty}(t)=\frac{1}{K}$ for any $t\leq T-t_0$. Therefore, the expected discounted cost can be rewrite as
\bee\label{equ_proof_epsi1}\begin{split} U(T)=&\sum_{t=0}^{T}\rho^t(A^2(t)+cp^2(t))\\
=&\sum_{t=0}^{T-t_0}\rho^t((A^{\infty}(t))^2+c(p^{\infty}(t))^2)\\&+\sum_{t=T-t_0+1}^{T}\rho^t(A^2(t)+cp^2(t))\\
=&\sum_{t=0}^{T-t_0}\rho^t((A^{\infty}(t))^2+c(p^{\infty}(t))^2)+\varepsilon_1(T). \end{split}\ene

Denote $\bar{\cU}=\max(A^2(t)+cp^2(t)|t\in[T-t_0+1,T])$ and $\underline{\cU}=\min(A^2(t)+cp^2(t)|t\in[T-t_0+1,T])$. Then, we have
\bee \varepsilon_1(T)\leq \sum_{t=T-t_0+1}^{T}\rho^t\bar{\cU}=\bar{\cU}\frac{\rho^{T-t_0+1}(1-\rho^{t_0})}{1-\rho}, \ene
and
\bee \varepsilon_1(T)\geq \sum_{t=T-t_0+1}^{T}\rho^t\underline{\cU}=\underline{\cU}\frac{\rho^{T-t_0+1}(1-\rho^{t_0})}{1-\rho}. \ene

As $T\rightarrow\infty$, we have $\bar{\cU}\frac{\rho^{T-t_0+1}(1-\rho^{t_0})}{1-\rho}=0$ and $\underline{\cU}\frac{\rho^{T-t_0+1}(1-\rho^{t_0})}{1-\rho}=0$. Thus, according to Squeeze Theorem, we have
\bee \lim_{T\rightarrow\infty}\varepsilon_1(T)=0. \ene

For finite horizon with steady state $Q, M$, we have \bee\begin{split}\label{equ_proof_epsi2} U^{\infty}(T)=&\sum_{t=0}^{T-t_0}\rho^t((A^{\infty}(t))^2+c(p^{\infty}(t))^2)\\
&+\sum_{t=T-t_0+1}^{T}\rho^t((A^{\infty}(t))^2+c(p^{\infty}(t))^2)\\
=&\sum_{t=0}^{T-t_0}\rho^t((A^{\infty}(t))^2+c(p^{\infty}(t))^2)+\varepsilon_2(T). \end{split}\ene

Similarly, we can show that $\lim_{T\rightarrow\infty}\varepsilon_2(T)=0$. Since $p(t)$ in (\ref{equ_pt_finite}) is the optimal price for finite horizon, we have $U(T)\leq U^{\infty}(T)$. Combine (\ref{equ_proof_epsi1}) and (\ref{equ_proof_epsi2}), we can see that for $\forall$ $T>0$, there always exists a $\varepsilon_T>0$ such that $U^{\infty}(T)\leq U(T)+\varepsilon_T$ with $\lim_{T\rightarrow\infty}\varepsilon_T=0$. \qed

As shown in Fig. \ref{fig_diffU}, the difference between $U^{\infty}(T)$ and $U(T)$ reduces as $T$ increases, which approaches $0$ for sufficiently large $T$. This is consistent with this proposition and tells that the simple pricing in (\ref{equ_pt_infinity_horizon}) performs well once $T$ is large (not necessarily infinite).

\section{Conclusion}\label{sec_conclusion}

In this paper, we have studies the dynamic pricing that minimizes the discounted AoI and payment over time. We have formulated this problem as a constrained nonlinear dynamic process under incomplete information about users' random arrival and private sampling costs. For analysis tractability, we linearize the nonlinear AoI evolution in the constrained dynamic programming problem by using the weighted time-average age to estimate the dynamic AoI reduction. It is shown that the estimator is appropriately designed to replicate the time-average term initially assumed. We further analyze the steady-state of the approximate dynamic pricing for infinite horizon and show that, as time goes to infinity, the approximate dynamic pricing can be further simplified to an $\varepsilon$-optimal version without recursive computing over time.

\bibliographystyle{IEEEtran}
\bibliography{AoIref}

\end{document}